\documentclass[11pt]{article}
\pdfoutput=1
\usepackage{jheppub}

\usepackage{epsfig}
\usepackage{latexsym}
\usepackage{amsfonts}
\usepackage{mathtools}
\usepackage{amsthm}
\usepackage{amssymb}
\usepackage{amsbsy}
\usepackage{multirow}
\usepackage{slashed}
\usepackage{amsmath,latexsym,amssymb}
\usepackage{graphicx}
\usepackage{psfrag}
\usepackage[latin1]{inputenc}
\usepackage[T1]{fontenc}
\usepackage{subfigure}
\usepackage{nicefrac}
\usepackage[stable]{footmisc}
\usepackage{rotating}
\usepackage{afterpage}
\usepackage{bbm}

\def\NO{\nonumber}

\newcommand{\be}{\begin{equation}}
\newcommand{\ee}{\end{equation}}

\def\bea{\begin{eqnarray}}
\def\eea{\end{eqnarray}}
\def\bi{\begin{itemize}}
\def\ei{\end{itemize}}
\def\bu{\begin{enumerate}}
\def\eu{\end{enumerate}}

\def\beqx{\begin{displaymath}}
\def\eeqx{\end{displaymath}}

\newcommand{\bmat}{\left(\begin{array}}
\newcommand{\emat}{\end{array}\right)}

\newcommand{\rf}[1]{(\ref{#1})}

\def\a{\alpha}

\def\c{\chi}

\def\e{\epsilon}
\def\f{\phi}
\def\g{\gamma}

\def\r{\rho}
\def\s{\sigma}
\def\t{\tau}

\def\bo{{\raise-.3ex\hbox{\large$\Box$}}}               
\def\face{{\raise.2ex\hbox{$\displaystyle \bigodot$}\mskip-2.2mu \llap {$\ddot
        \smile$}}}                                   
\def\>{\rangle}                                      
\def\<{\langle}                                      

\def\leftrightarrowfill{$\mathsurround=0pt \mathord\leftarrow \mkern-6mu
        \cleaders\hbox{$\mkern-2mu \mathord- \mkern-2mu$}\hfill
        \mkern-6mu \mathord\rightarrow$}        
\def\dvec#1{\vbox{\ialign{##\crcr
        \leftrightarrowfill\crcr\noalign{\kern-1pt\nointerlineskip}
        $\hfil\displaystyle{#1}\hfil$\crcr}}}           

\RequirePackage[normalem]{ulem} 
\RequirePackage{color}\definecolor{RED}{rgb}{1,0,0}\definecolor{BLUE}{rgb}{0,0,1} 
\definecolor{purple}{rgb}{0.78,0.18,0.77}

\def\-{\hphantom{-}}

\title{\rm {\bf \LARGE Journeys Through Antigravity?}}
\author{John Joseph M. Carrasco,} 
\author{Wissam Chemissany, and} 
\author{Renata Kallosh}

\affiliation{Stanford Institute for Theoretical Physics and Department of Physics, Stanford University, 
382 Via Pueblo Mall, Stanford, CA 94305-4060, U.S.A. }

\emailAdd{jjmc@stanford.edu} 
\emailAdd{wissam@stanford.edu} 
\emailAdd{kallosh@stanford.edu}

\abstract{
A possibility of  journeys through antigravity has recently been proposed, with the suggestion that Weyl-invariant extension of  scalars coupled to Einstein gravity allows for an unambiguous classical evolution through cosmological singularities in anisotropic spacetimes. We compute the Weyl invariant  curvature squared and find that it blows up for the proposed anisotropic solution both at the Big Crunch  as well as at the Big Bang. Therefore the cosmological singularities are not resolved by uplifting Einstein theory to a Weyl invariant model.

}

\begin{document}  \maketitle

\section{Introduction}

Singularities in static solutions of general relativity are known to be of two types.  The first, known as {\it coordinate singularities}, have the property of being absent in some choice of coordinates.   The second, known as {\it true singularities}, have the property of being unavoidable for any choice of coordinates.  For example in the Schwarzschild black hole solution there are two singularities: one at the center $r=0$, and one at the event horizon $r=r_h$.  Note that the value of the scalar $R_{\mu\nu\lambda \delta}R^{\mu\nu\lambda \delta}$ is the same in all coordinate systems and therefore it is viewed in general relativity as a  {\it physical observable}.   The curvature squared for the black hole spacetime is coordinate independent and given by 
\be
R_{\mu\nu\lambda \delta}R^{\mu\nu\lambda \delta}={48 M^2 \over r^6}  \ .
\ee
Thus, as the singularity at $r=r_h$ is not present in the curvature invariant, it may be removed by the change of coordinates, which is exactly what takes place when Kruskal-Szekeres coordinates are used. However, at $r=0$ the black hole solution has a true intrinsic singularity, visible to $R_{\mu\nu\lambda \delta}R^{\mu\nu\lambda \delta}$.

It was pointed  out in \cite{Linde:1979kf} that models describing massless fields including scalar fields with conformal coupling to gravity in a uniform and isotropic Friedmann universe, could allow one to continuously cross the boundary between gravity and antigravity. However, it was soon realized that if the universe is not absolutely isotropic, the tiny deviations from isotropy become amplified, leading to a true cosmological singularity at the boundary between  gravity and antigravity regimes~\cite{Starobinskii1981}. This conclusion was verified, confirmed and generalized, see for example,~\cite{Futamase:1989hb, Abramo:2002rn}. One may come to a similar conclusion by studying also slightly inhomogeneous spaces~\cite{Starobinskii1981, Caputa:2013mfa}. Qualitative reasons for such behavior are easy to understand. In the vicinity of the transition from gravity to antigravity, the effective Planck mass squared disappears, the effective gravitational constant blows up, which leads to an infinitely growing gravitational instability and the development of a true cosmological singularity at the transition point.

In a closely related context, it has been suggested recently in \cite{Bars:2011th,Bars:2011aa,Bars:2012fq} that uplifting general relativity via a Weyl-invariant extension allows the definition of a classical evolution through cosmological singularities. For this to happen, the geodesically complete space-time  must include some domains of antigravity. This would contradict the results of \cite{Starobinskii1981, Futamase:1989hb, Abramo:2002rn, Caputa:2013mfa}, but it was claimed in \cite{Bars:2011th,Bars:2011aa,Bars:2012fq} that the Weyl-invariant extension renders the standard cosmological singularity unphysical, just like the transition to the Kruskal-Szekeres coordinates alleviates the event horizon singularity for black holes, which is present in the Schwarzschild coordinates.

However, in their investigation of the cosmological singularity, the authors of \cite{Bars:2011th,Bars:2011aa,Bars:2012fq} did not  study any Weyl-invariant extensions of such curvature invariants as $R_{\mu\nu\lambda \delta}R^{\mu\nu\lambda \delta}$ or  $C_{\mu\nu\lambda \delta}C^{\mu\nu\lambda \delta}$, which blow up at the cosmological singularity in the Einstein frame. Instead of that, they studied only one specific invariant  which was non-singular in any frame, including the Einstein frame. Therefore investigation of the transition to the antigravity regime and back in \cite{Bars:2011th,Bars:2011aa,Bars:2012fq} was incomplete. 
For locally conformally invariant theories, ref.~\cite{KL} suggests natural Weyl curvature invariants which generalize the $R_{\mu\nu\lambda \delta}R^{\mu\nu\lambda \delta}$ invariant to Weyl symmetry.  These should play the role of  physical observables whose properties are independent on the choice of a Weyl gauge and of a coordinate system. The simplest possible such invariant is
\be
I= f(X, \bar X) C_{\mu\nu\lambda \delta}C^{\mu\nu\lambda \delta} \ .
\label{I}
\ee
Here $C^\mu{}_{\nu\lambda\delta}$ is the  Weyl curvature tensor of a conformal weight $w=0$, the traceless part of the Riemann tensor, and the function of scalars $f(X, \bar X)$ has a  weight $w=-4$.

These are natural observables to use when evaluating the claims of refs.~\cite{Bars:2011th,Bars:2011aa,Bars:2012fq} that the geodesic  incompleteness of cosmological  solutions of the Einstein gravity can be removed by a better choice of the conformal gauge associated with the Weyl symmetry. The arguments put forth were based on their construction of a variable which happens to be  Weyl-invariant and analytic at generic cosmological singularities. First we will clarify certain aspects of the near singularity solution proposed in refs.~\cite{Bars:2011th,Bars:2011aa,Bars:2012fq}. We will argue that the solution there, {\it presented} as  Bianchi IX, actually becomes the Bianchi I solution near singularities in the approximation neglecting spacial curvature.
After reviewing and clarifying their framework, we will consider the relevant observable -- the Weyl curvature invariant \rf{I} on solutions near cosmological singularities, at the Big Crunch when gravity turns into antigravity and at the Big Bang where antigravity turns into gravity.  We will find that the Weyl invariant curvature is gauge independent and infinite both at the Big Crunch as well as at the Big Bang.  


\section{A proposal for a journey through antigravity \cite{Bars:2011th,Bars:2011aa,Bars:2012fq}}
The proposal is to use Weyl symmetry to solve cosmological equations and to traverse cosmological singularities. The starting point  is a Lagrangian which depends on two conformally coupled scalars. \begin{equation}
\mathcal{L}\left(  x\right)  =\frac{1}{12}\left(  \phi^{2}-s^{2}\right)
R\left(  g\right)  +\frac{1}{2} \left(  \partial_{\mu}\phi
\partial_{\nu}\phi-\partial_{\mu}s\partial_{\nu}s\right) g^{\mu\nu} -\phi
^{4}f\left(  s/\phi\right)  \ . \label{action1}%
\end{equation}
Also some terms with radiation and matter can be added. Note that the scalar $\phi$ has a `wrong sign' kinetic term, however, the action has a Weyl symmetry, which means that there are no ghosts, as one can see in  the equivalent form of the action below in\rf{Einst}. 
The action 
\be
S=\int d^4 x \sqrt{-g} \, \mathcal{L}
\ee
is Weyl invariant  under Weyl transformations
\be
\phi\rightarrow \Omega \phi \, , \qquad s \rightarrow \Omega s \, , \qquad g_{\mu\nu} \rightarrow \Omega^{-2} g_{\mu\nu}  \ .
\label{Weyl}\ee
This means that the scalars have conformal weight $w=1$ and the metric $g_{\mu\nu} $ has $w=-2$, $\mathcal{L}$ has weight $w=4$ and $\sqrt{-g}$ has $w=-4$, where 
$ g\equiv  \det \, g_{\mu\nu}
$.
One may gauge-fix Weyl symmetry to recover {\it the Einstein frame general relativity }
\be
 (\phi^{2}-s^{2})_E= 6
\ee
in units $M_{Pl}=1$. The action becomes
\be
S_E=\int d^4 x \sqrt{-g}_E \Big [   \frac{1}{2}
R\left(  g\right)_E  -  \frac{1}{2}\partial_{\mu}\sigma
\partial_{\nu}\sigma g^{\mu\nu}_E -V_E(\sigma) \Big ]  \ ,
\label{Einst}\ee
where $\tanh {\sigma\over \sqrt 6} = {s\over \phi}$. Obviously, there are no ghosts in this model.

Note the  significant difference between Weyl symmetric models  without ghosts and the `nonsingular bounce' type models studied in
\cite{Allen:2004vz, Barrow:2013qfa} where
\begin{equation}
\mathcal{L}\left(  x\right)  =\frac{1}{2}
R\left(  g\right)  +\frac{1}{2} \left(  \partial_{\mu}\phi
\partial_{\nu}\phi-\partial_{\mu}s\partial_{\nu}s\right) g^{\mu\nu} - V(s) \ .
   \label{actionGhost}%
\end{equation}
Such actions do not have Weyl symmetry. Their scalar $\phi$ with the `wrong sign' kinetic term is a ghost field  required for  non-singular bounce models.

Returning to Weyl invariant no-ghosts models, we may be interested in various gauge fixing conditions which break Weyl symmetry.
One of the gauges fixing Weyl symmetry is given by the condition that  
\be
-g\equiv  - (\det \, g_{\mu\nu})=1 \ .
\ee
This is a condition introduced in studies of general relativity in \cite{vanderBij:1981ym}. In the context of Weyl invariant theories it is important that $g$ has a conformal weight $w= -4$ which is the reason why imposing this condition means breaking  Weyl symmetry. It was  called {\it $\gamma$-gauge}  in \cite{Bars:2011th}, $-g_\gamma= - (\det \, g_{\mu\nu})_\gamma=1$.  It was  proposed that this gauge  makes visible a path through antigravity avoiding cosmological singularities of the Einstein theory. In this gauge the action \rf{action1} is 
\be
S=\int d^4 x \Big [\frac{1}{12}\left(  \phi^{2}-s^{2}\right)
R\left(  g\right)_\gamma  +\frac{1}{2} \left(  \partial_{\mu}\phi
\partial_{\nu}\phi-\partial_{\mu}s\partial_{\nu}s\right) g^{\mu\nu}_\gamma -V_\gamma(s, 
\phi)\Big]\,.
\ee

We may solve the classical equations of motion following from the Weyl invariant action or from the ones where the Weyl symmetry was gauge-fixed. The solutions in different gauges will be related to each other by some Weyl transformation of the form given in  \rf{Weyl}.

We are interested here only in solutions where the metric has definite anisotropy as isotropic solutions do not suffer from cosmological singularities when approaching the transition from gravity to antigravity~\cite{Linde:1979kf}.
Thus following  \cite{Bars:2011aa} we focus our attention on an anisotropic Bianchi IX  metric which describes a mixmaster universe \cite{Misner:1969hg}.
\be ds^{2}=a^{2}(\t)\left(-d\t^{2}+ds_{3}^{2}\right).\ee

The Bianchi IX model describes a homogenous anisotropic 3-dimensional geometry with $S^{3}$ topology and isometry group $SU(2)$. In notation of \cite{Misner:1969hg}
\be
ds_{3}^{2} = \sum_{i=1}^3 (l_i  \sigma_i)^2
\label{d3}\ee
The corresponding metric is constructed 
using the left-invariant 1-forms $\s_{i}$,  
\bea  
\s_{1}=\sin \psi \sin \theta d\f+\cos \psi d\theta,\quad 
 \s_{2}= \cos \psi \sin \theta d\f-\sin \psi d\theta,\quad
\s_{3}=\cos \theta d\f+d\psi,
\eea 
\bea   l_{1}^2=r^2 e^{ \left(\sqrt{2/3} \, \a_{1}+\sqrt{2}\,  \a_{2}\right)}, \quad l_{2}^2=r^2 e^{ \left(\sqrt{2/3} \, \a_{1}-\sqrt{2}\,  \a_{2}\right)}, \quad l_{3}^2=r^2 e^{- \sqrt{8/3} \, \a_{1}}\, .
\label{forms} \eea
 where we have introduced explicitly the constant  ``radius of curvature'' $r$, as defined in \cite{Landau:1989gn}. The functions $\alpha_1(\tau), \alpha_2(\tau)$ need depend only on the time $\tau$~\cite{Bars:2011aa}.

In Bianchi IX models the spatial-curvature is anisotropic,  the 3-curvature of the Bianchi IX model is generically non-vanishing, and only becomes positive when the dynamics approach isotropy\footnote{ The case of non-vanishing 3-curvature for isotropic FLRW consideration has been studied in ref.~\cite{Bars:2012mt}. }. For the explicit expressions of the curvature components and the corresponding analysis we refer to ref.~\cite{Barrow:2013qfa}.
For the proposed antigravity solution in refs.~\cite{Bars:2011th,Bars:2011aa,Bars:2012fq}, they only consider  the situation where the  curvature of the Bianchi IX is neglected\footnote{We are grateful to John Barrow for emphasizing to us that generically the 3-curvature cannot be neglected near singularities.}.  As we will now explain, this actually brings the ansatz for the metric to a Bianchi I and the anisotropic solution with vanishing spatial curvature as we will discuss.   
  
Therefore, we first will inspect more carefully  the meaning of the statement in \cite{Bars:2012fq} that the spatial curvature of Bianchi IX is neglected near the cosmological singularity.  One may rightly question whether this is actually valid.  The time-dependent curvature anisotropy is always important and drives the chaotic oscillations~\cite{Belinsky:1982pk} of the scale factors near the singularity. There is an infinite number of such oscillations on any open interval around $t =0$. However in presence of  a massless scalar field
it might be possible to find a solution without such chaotic oscillations, allowing the neglect of curvature as in ref.~\cite{Bars:2012fq}.  The model in \cite{Bars:2011th,Bars:2011aa,Bars:2012fq} in fact has a  scalar field, see \rf{Einst} and since the potential is neglected near singularity it makes the scalar massless.

 In notation of \rf{d3}-\rf{forms} the vanishing  Bianchi IX curvature means that the radius of the curvature $r$ goes to  $r\rightarrow \infty$ and the space tends to a flat one. The rescaled by the radius of curvature $r$  1-forms $\tilde \sigma_i= r \s_{i}$ satisfy the following rescaled $SU(2)$ relations
\be d\tilde  \s^{i}+\frac{1}{2 r}\e^{ijk}\,  \tilde \s_{i}\wedge \tilde \s_{k}=0,
\label{SU2}\ee
Near the cosmological singularities terms involving the spatial curvature becomes negligible according to \cite{Bars:2011th,Bars:2011aa,Bars:2012fq}. In this approximation one finds that eq. \rf{SU2} degenerates and the forms 
$\tilde  \s_{i}$ become closed:
\be
d\tilde  \s^{i}|_{r\rightarrow \infty}  \rightarrow 0 \qquad \Rightarrow \qquad \tilde  \s^{i}= d x^i \ .
\ee
This means that near the singularities the metric ansatz before fixing any gauge becomes
\be\label{metric} 
ds^{2}=a^{2}(\t)\left(-d\t^{2}+e^{ \left(\sqrt{2/3} \, \a_{1}+\sqrt{2}\,  \a_{2}\right)} dx^{2}+ e^{ \left(\sqrt{2/3} \, \a_{1}-\sqrt{2}\,  \a_{2}\right)} dy^2 + e^{- \sqrt{8/3} \, \a_{1}} dz^2\right).\ee
We will confirm in the next section that the actual anisotropic solution of equations of motion presented in \cite{Bars:2011th,Bars:2011aa,Bars:2012fq} is valid for the Bianchi I case.

By looking at this metric one may think that in a Weyl gauge where 
$
a^{2}(\t)_\gamma=1$
 it might be possible to avoid the cosmological singularity which in other gauges is related to the scale factor of the universe collapsing to zero. 
  We will check if this expectation is indeed realized. 


\section{An analytic Weyl invariant}
It was observed in \cite{Bars:2011th,Bars:2011aa,Bars:2012fq} that the model has a Weyl invariant given by the following expression
\be
\c(\t)\equiv (-g)^{1\over 4}\  {\phi^{2}-s^{2}\over 6}  \ .
\label{WI}\ee
Under Weyl transformations $\phi^{2}-s^{2}$ has $w=2$ and $(-g)^{1\over 4}$ has $w=-2$. It will be therefore useful to extract the value of this Weyl invariant in various Weyl gauges to double check that the solutions are correct and most of all to make sure that near the singularities the invariant does not blow up. Note, however, that under the change of coordinates $\c$ is not a scalar, the square of it,  $\c^2= (-g)^{1\over 2}\Big({\phi^{2}-s^{2}\over 6} \Big)^2$,  is a scalar density. 

The $\gamma$-gauge effective action given in ref.~\cite{Bars:2011aa}, which we quote here:
\begin{equation}\label{theirSe}
S_{\gamma}^{\,\rm eff}  =   \int d\tau\left( 
\frac{1}{2e}\left[ -\dot{\phi}_{\gamma}^2  +  \dot{s}_{\gamma}^2 + \frac{\kappa^2}{6}(\phi_{\gamma}^2 - s_{\gamma}^2)(\dot{\alpha}_1^2+ \dot{\alpha}_2^2)\right]  -e \rho_r  \right)\,
\end{equation}
 shows that the solution was obtained for the Bianchi I metric, as we will explain.  In general, in this gauge with $a=1$ we would expect the curvature part of the effective action to be of the form~\cite{KL},
 \begin{equation}
 \label{renAction}
-{1\over 6} \sqrt {\det g} \, {\cal N}(X, \bar X) \left[ \alpha \gamma^{1/2} (K_{ij} K^{ij} - K^{(2)} + \, {}^{(3)}R - 2\gamma^{1/2} ( K)_{, 0}+(K\beta^i - \gamma^{ij} \alpha_{,j})_{, j}\right].
\end{equation}
  We may compare eq.~(\ref{theirSe})  with the general formula eq.~(\ref{renAction}) and we can see that the term with ${}^{(3)}R$ is absent from the former.  The expression for the ${}^{(3)}R$ for Bianchi IX can be found, e.g. in ref.~\cite{Barrow:2013qfa}.  See also eqs.~(8-9) in  \cite{Bars:2012fq} where the analogous expressions were given and the curvature parameter $k\to 0$, making the solution consistent with Bianchi I.  
 
When equations of motion are solved the Weyl invariant $\c(\t)$ turns out to be given by the following expression
\be
\c(\t)=2 \bar{\t}(p+\r_{r}\bar{\t})\, , \qquad \bar{\t}\equiv \frac{ \t}{\sqrt{6}} \ ,
\ee
where $\rho_{r}$ is a constant responsible for radiation, which is added to the actions above.
This invariant $\c(\t)$  vanishes at 
\be \t= \bar{\t}=0,\quad \textrm{and}\,\, \, \t_c = \sqrt{6} \, \bar{\t}_c   \qquad \bar{\t}_{c}=- \frac{p}{\r_{r}}.\ee
The complete solution depends on scalars functions which have conformal weight $w=0$ and are therefore Weyl gauge-independent. These are
\bea && \s(\t)= \frac{\sqrt{6} p_{\s}}{2  p}\ln\left|\frac{\bar{\t}}{T (p+\r_{r}\bar{\t})}\right|,\\
&&\a_{1}(\t)=\frac{\sqrt{6} p_{1}}{2  p}\ln\left|\frac{\bar{\t}}{T_{1} (p+\r_{r}\bar{\t})}\right|,\label{a1}\\
&& \a_{2}(\t)=\frac{\sqrt{6} p_{2}}{2  p}\ln\left|\frac{\bar{\t}}{T_{2} (p+\r_{r}\bar{\t})}\right| \label{a2}.\eea
Here
$p, p_\sigma, p_1, p_2$ are constants since the potential and the spatial curvature are ignored near the singularities and
$ 
p=\sqrt{p^{2}_{\s}+p_{1}^{2}+p_{2}^{2}}$.
Since the solutions were obtained by solving  differential equations, there are integration constants there, $T, T_1, T_2$. There are no constraints on all these constants apart from the fact that we exclude the case $p_1=p_2=0$ which would make the anisotropy function $\alpha_1, \alpha_2$ time-independent, since they satisfy the following equations
\be \dot{\a}_{1}=\frac{p_{1}}{2 p} \frac{1}{\bar{\t}}, \qquad  \dot{\a}_{2}=\frac{p_{2}}{2 p} \frac{1}{\bar{\t}}.\ee 
The solutions for the scalars $\phi$ and $s$ which have $w=1$ clearly depend on the choice of the conformal gauge. For example, in the $\gamma$ gauge they are given by the following expressions
\bea && \frac{1}{\sqrt{6}}(\f_{\g}+s_{\g})=\sqrt{T}(p+\rho_{r} \bar{\t})\left|\frac{\bar{\t}}{T(p+\rho_{r} \bar{\t})}\right|^{(p+p_{\s})/2p},\nonumber \\
\nonumber\\
&& \frac{1}{\sqrt{6}}(\f_{\g}-s_{\g})=\frac{2\bar{\t}}{\sqrt{T}}\left|\frac{\bar{\t}}{T(p+\rho_{r} \bar{\t})}\right|^{-(p+p_{\s})/2p} \ . \eea
This confirms that,    with $g_\gamma=a(\t)_\gamma=1$, one finds that
$
\c(\t)_\gamma=2 \bar{\t}(p+\r_{r}\bar{\t})\, . 
$
In the Einstein gauge, where ${\phi^{2}-s^{2}\over 6}|_E =1 $ one finds that 
\be
a^2(\t)_E= \c(\t)= 2 \bar{\t}(p+\r_{r}\bar{\t})\, , 
\ee
in agreement with the fact that $\c(\t)$ is Weyl invariant.

\subsection{Cosmological singularities and the proposed recipe for traversing via antigravity}

In the Einstein frame the vanishing of the scale factor $a^2(\t)_E$ takes place at two points in time, at Big Bang at $\t_{BB}=0$ and at Big Crunch at $\t_{BC}= -\sqrt 6 \, p/\rho_r$
\be 
a^2(\t)_E= 0 \qquad \Rightarrow \qquad 
\t_{BB}= \bar{\t}=0,\quad \textrm{and}\,\ \, \t_{BC}= \sqrt{6} \, \bar{\t}_c \, ,  \qquad \bar{\t}_{c}=- \frac{p}{\r_{r}}\ee
This is qualified in \cite{Bars:2011th} as a geodesic incompleteness of the model in the Einstein frame, where it is also suggested that things may become better in the gauge where $a^2(\t)_\gamma=1$.

It is proposed that in $a^2(\t)_\gamma=1$ gauge there is a resolution of the cosmological singularities which are present in the Einstein frame. The idea of traversing cosmological singularities and of a complete journey through spacetimes including antigravity is summarized in \cite{Bars:2012fq}. In this gauge the scalar fields are dynamical, the value of the expression $\frac{1}{12}\left(  \phi^{2}-s^{2}\right)$ in front of $R$ in \rf{action1} may take various values, positive and negative. Restoring the units with Newton gravitational coupling $G_N$ we have, in general that 
\be
{\sqrt {-g } \over 16 \pi G_N}  \frac{\phi^{2}-s^{2}}{6} R= {\sqrt {-g } \over 16 \pi \tilde G_N}    R \ ,
\ee
where for example in $\gamma$-gauge
\be
\tilde G_N (\t)|_\gamma = \frac {6 \, G_N }{\phi_\gamma(\t)^{2}-s_\gamma(\t)^{2}} \ .
\ee
In $\gamma$-gauge where $\sqrt {-g }=1$ and $\phi_\gamma$ and $s_\gamma$ are dynamical, one finds that vanishing of $\c(\t)_\gamma= {\phi_\gamma^{2}-s_\gamma^{2}\over 6}$ has an interpretation of switching the sign of the effective gravitational coupling $\tilde G_N (\t)$ from gravity to antigravity, when $\c(\t)_\gamma$ is vanishing. It also means that at this points in time, at $\t_{BB}= 0$ and at $\t_{BC}= -\sqrt 6 \, p/\rho_r$ there is a gravity/antigravity switch.

According to \cite{Bars:2011th,Bars:2011aa,Bars:2012fq} the resolution of cosmological Big Bang and Big Crunch singularities takes place in such a `Weyl uplifted' model since the `failure of the geometry in the Einstein gauge does not imply the failure of the geometry in other Weyl gauges'. In particular, in the $\gamma$-gauge one can perform an analytic continuation through the singularity. The reason is that near both of these singularities the Weyl invariant $\c(\t)$ is analytic. 

The {\it recipe} for traversing the singularities is  first to enter antigravity at $\t_{BC}= -\sqrt 6 \, p/\rho_r$, spend some time in the antigravity spacetime until $\t=0$, and return back to gravity for $\t>0$. One  would then define the scale factor of the universe in the Einstein frame as follows 
\be
a^2_E(\t)= |\c(\t)|= |2 \bar{\t}(p+\r_{r}\bar{\t})| \ .
\ee
The fact that it vanishes a couple of times, at the Big Crunch at $\t_{BC}= -\sqrt 6 \, p/\rho_r$ and at the Big Bang at $\t=0$ looks harmless:  the curvature $R_{\mu\nu\lambda\delta}(g_E)$ transforms under Weyl transformations, so it was argued there is no need to worry about it as there exists the Weyl invariant \rf{WI} which is analytic at $\t_{BC}= -\sqrt 6 \, p/\rho_r$ and at $\t=0$.

However, naturally (see e.g. \cite{KL}) there are other Weyl invariants depending on the Weyl curvature tensor which may be multiplied by some function of scalars so that such invariants have $w=0$ and do not depend on the choice of the conformal gauge. The simplest one was shown in \rf{I} where one can take any function of $\phi$ or $s$ with $w=2$ to balance the $w=4$ from $C^2$.

\section{Weyl invariant curvature squared blows up}
An example of the Weyl invariant curvature squared  is given by
\be
I(x) = \left( {\phi^{2}-s^{2}\over 6}  \right)^{-2} C_{\mu\nu\lambda\delta} C^{\mu\nu\lambda\delta} \ .
\label{curv}\ee
In $\gamma$-gauge where  $- g(\t)=1$,  
\be  \left( {\phi^{2}_{\g}-s^{2}_{\g}\over 6}  \right)^{-2}=  \, \c_\gamma(\t)^{-2}= (2 \bar{\t}(p+\r_{r}\bar{\t}))^{-2} \ ,
\ee
 and the metric is $ds^{2}_\gamma= -d\t^{2} + ds_{3}^{2}$ where
\be\label{metricG} 
 ds_{3}^{2} = e^{ \left(\sqrt{2/3} \, \a_{1}+\sqrt{2}\,  \a_{2}\right)} dx^{2}+ e^{ \left(\sqrt{2/3} \, \a_{1}-\sqrt{2}\,  \a_{2}\right)} dy^2 + e^{- \sqrt{8/3} \, \a_{1}} dz^2.\ee 
The explicit expressions for anisotropies $\alpha_1(\t)$ and $\alpha_2(\t)$ are given in
 \rf{a1}, \rf{a2}
We find the following expression for the  Weyl invariant curvature squared in the $\gamma$-gauge
\bea
 I&=&\frac{ 3}{4 \bar{\t}^6 (p+\r_{r} \bar{\t})^6}\Big[p^2 \left(p_{1}^2+p_{2}^2\right)+4 p \r_{r}  \left(p_{1}^2+p_{2}^2\right)\bar{\t}+2 p p_{1}\left(p_{1}^2-3 p_{2}^2\right)\NO\\&+&4 \r_{r}^2 \bar{\t}^2
   \left(p_{1}^2+p_{2}^2\right)+4p_{1} \r_{r} \bar{\t} \left(p_{1}^2-3 p_{2}^2\right)+\left(p_{1}^2+p_{2}^2\right)^2\Big] .
   \eea
 In Einstein gauge, where $\left( {\phi^{2}-s^{2}\over 6}  \right)^{-2}=1 $ and 
$ ds^{2}_E= a^{2}_E(\t)\left(-d\t^{2}+ds_{3}^{2}\right)$
and  $a^2_E(\t)= \c(\t)= 2 \bar{\t}(p+\r_{r}\bar{\t})$,
we find for $I_E= C^{2}(g_E)$
as expected the same value of the invariant, 
\be
 I_E= I_\gamma \ .
\ee
 In fact, the computation of the Weyl curvature squared in anisotropic Bianchi I models in the Einstein frame was already performed in \cite{Gron:2000dp} and the result is compatible with ours.

 We may now proceed to singularities.
  At the Big Crunch near  $\tau_{BC}= -\sqrt 6\, {p\over \rho_r}$ the Weyl invariant curvature squared blows up 
\be I|_{\tau \rightarrow \t_{BC} } \rightarrow \frac{  2\cdot 3^4\left[p^2 \left(p_{1}^2+p_{2}^2\right)-2 p p_{1}\left(p_{1}^2-3 p_{2}^2\right)+\left(p_{1}^2+p_{2}^2\right)^2\right]}{ p^6(\t- \t_{BC})^6}\rightarrow\infty \ .
\ee
At the  Big Bang at $\tau \rightarrow 0$    we also find 
   \be I|_{\tau \rightarrow 0} \rightarrow \frac{2\cdot 3^4  \left[p^2 \left(p_{1}^2+p_{2}^2\right)+2 p p_{1}\left(p_{1}^2-3 p_{2}^2\right)+\left(p_{1}^2+p_{2}^2\right)^2\right]}{ \,  p^6 {\t}^6} \rightarrow\infty \ .
   \ee  
 Note that our main condition was that $p_1$ and $p_2$ do not vanish, i.e. there is a time dependent anisotropy in $\alpha_i(\t)$. Under these restrictions, for generic `anisotropy velocities' $p_1$  and $p_2$ the  Big Crunch and Big Bang singularities of the Weyl  invariant curvature squared are unavoidable. Note that the combination of these two expressions is strictly positive for non-vanishing arbitrary $p_i$, namely
\be
(I|_{\tau \rightarrow 0} ) p^6 \t^6 + (I|_{\tau \rightarrow \t_{BC} }) p^6 (\t- \t_{BC})^6 \quad \rightarrow \quad p^2 \left(p_{1}^2+p_{2}^2\right)+\left(p_{1}^2+p_{2}^2\right)^2 >0 \ .
\ee
Therefore there is no way to avoid both singularities in anisotropic solutions.

 Thus, the Weyl invariant curvature \rf{curv} at the Big Bang and  at the Big Crunch singularity is divergent for solutions of classical equations of motion in \cite{Bars:2011th,Bars:2011aa,Bars:2012fq}, confirming that it is a `true singularity'. \section{Discussion}

Here we consider the solutions of Einstein equations suggested in refs.~\cite{Bars:2011th,Bars:2011aa,Bars:2012fq}, claimed to allow the transition through antigravity, and  put forth as resolving the cosmological singularities. Computing the value of the Weyl invariant curvature squared we find, rather, clear divergences at cosmological singularities: there is no resolution and no journeys through spacetime including antigravity under the condition that there is some time-dependent anisotropy.

Is this obvious or does it depend solely of specifics of this particular solution where we were able to compute the invariant curvature in all Weyl gauges?  We argue it is obvious, based on the structure of the the Weyl invariant curvature. In the Einstein frame $\phi^{2}-s^{2}=6$

\be 
I(x) = \left( {\phi^{2}-s^{2}\over 6}  \right)^{-2} C_{\mu\nu\lambda\delta} C^{\mu\nu\lambda\delta}= C_{\mu\nu\lambda\delta} C^{\mu\nu\lambda\delta}(g_E) \ .
\label{Inv}\ee
The square of the Weyl tensor general relativity is given by the following combination
\be
C_{\mu\nu\lambda\delta} C^{\mu\nu\lambda\delta}(g_E)= R_{\mu\nu\lambda\delta} R^{\mu\nu\lambda\delta}(g_E)- 2 R_{\mu\nu} R^{\mu\nu}(g_E)+ {1\over 3} R^2(g_E) \ .
\label{C2}\ee
 Since $R_{\mu\nu}$ is proportional to the energy-momentum tensor $T_{\mu\nu}$, we find that failing miraculous cancellations rendering Einstein frame $C_{\mu\nu\lambda\delta} C^{\mu\nu\lambda\delta}(g_E)$  finite, the usual cosmological singularities in the Einstein frame due to  
$R_{\mu\nu\lambda\delta} R^{\mu\nu\lambda\delta}(g_E)$ and/or  $T_{\mu\nu} T^{\mu\nu}$ and/or $T^2$ being infinite, will also be present in any other Weyl geometry. 

In the particular example  \cite{Bars:2011th,Bars:2011aa,Bars:2012fq} which we studied here, the antigravity regime in the Weyl uplifted geometry with $a(t)=1$ is separated from the gravity regime by an infinite Weyl invariant curvature singularity. In standard Einstein geometry at the Big Bang or Big Crunch we just observe that the combination of the squares of Riemann curvature,  Ricci,  and scalar curvature in \rf{C2} is infinite: thus we have the standard textbook classical cosmological 
singularity \cite{Landau:1989gn}. In Weyl uplifted geometry with $a(t)=1$ the singularity has an additional interpretation of the point where gravity would turn into antigravity. However, since we observe that the Weyl invariant curvature is infinite for solutions with anisotropy, the universe cannot pass from gravity to antigravity and back. This statement is valid independent of the choice of the conformal gauge as $I(x)$ in \rf{Inv} is gauge independent.

\section*{Acknowledgments}

We would like to thank S. Kachru and A. Linde for stimulating discussions. We are also very appreciative for clarifying correspondence with J.~Barrow, S.~Deser, and V. Mukhanov. We are happy to acknowledge the use of xCoba~\cite{xCoba}, part of the xAct free collection of Mathematica packages for tensor computer algebra.  JJMC, WC and RK are supported by Stanford Institute for Theoretical Physics (SITP), and  the NSF Grant PHY-1316699.  JJMC and RK are also supported by  the John Templeton foundation grant `Quantum Gravity Frontiers'.  WC is supported by the ArabFund for Economic and Social Development. 

\

\

\noindent {{\bf Note added:} }  

\

After this paper was submitted, Bars, Steinhardt, and Turok (BST) agreed with our statement that  the Weyl invariant curvature squared constructed in \cite{KL} is singular on their solutions \cite{Bars:2011th,Bars:2011aa,Bars:2012fq}, but claimed that it does not change their conclusions  \cite{Bars:2013qna}. They also suggest that certain massive geodesics defined in the classical background may ``continue'' through the Big Crunch and the Big Bang. We are grateful to the referee for the suggestion to respond in this added note.

The presence of Weyl invariant curvature singularities at finite points in time renders the ongoing discussion in refs.~\cite{Bars:2011th,Bars:2011aa,Bars:2012fq,Bars:2013qna}  highly problematic.   Physical infinities mean the classical theory is no longer in an applicable regime, and any associated discussion of geodesics in a classical background cannot be supported.  These infinities  manifest the need for the underlying UV-complete quantum theory of gravity with  the old recognition that when energy scales get high enough, curvature great enough, quantum mechanics has to resolve where classical general relativity is no-longer predictive.  Ref.~\cite{Bars:2013qna} explicitly dismisses the curvature singularities, lumping them amongst  infinitely-many other Weyl-invariant diverging quantities.  We will first make the case of physical relevance of the curvature singularities, then discuss the necessity of UV completion for this program, rendering all such discussions of classical geodesics irrelevant.  

Curvature singularities at finite points in time are associated with breakdown of GR theories for very physical reasons -- they are intimately related to the tidal forces experienced by bodies.   To sail through a singularity one cannot simply match particular massive geodesics and claim predictive traversal.   Within the classical regime, one should also consider null-geodesics, the geodesics of photons and gravitons, and study whether the tidal forces, stresses, and energies remain bound below the UV-cutoff (typically taken to be the Planckian scale).    It is worth considering a  closely related study of the cosmological singularity of the Kasner solution~\cite{Belinski:2001ph}, where it was demonstrated how the infinite tidal forces of null-congruence relate to the infinite curvature quadratic and cubic invariants.     It is hard to understand in what physical sense something might ``traverse'' paths buffeted by infinite tidal forces, independent of how ``smoothly'' they may be sailing from a particular point of view.

A classical GR analysis of the Raychaudhuri's type equations, analogous to the one in \cite{Belinski:2001ph}, was not performed in refs.~\cite{Bars:2011th,Bars:2011aa,Bars:2012fq,Bars:2013qna}. However, we believe that our results make such investigation unnecessary.
  The point is that the infinite curvature means the classical theory has broken down and the modeled system is in a regime where the classical methods are no longer applicable.
All calculations informed by densities/curvatures/energies above the UV-cutoff can not be trusted, which is one of the many reasons why the standard criterion for the existence of the cosmological singularity is the singular behavior of geometric invariants, such as the squared curvature tensor or the square of the Weyl tensor, or higher order invariants, see e.g. discussions of this issue in  \cite{Landau:1989gn,Wald:1984rg} and in \cite{KL}. Indeed, as soon as the universe enters the vicinity of the Big Crunch or the Big Bang where the curvature invariants exceed their Planck values, one can no longer trust any results based on solutions of the classical Einstein equations, which makes any claims about sailing through the singularity unreliable.

One could  have hoped that this problem may be alleviated by finding some Weyl gauge where the curvature invariants are finite. However, as we already emphasized in this paper, as well one of the authors in \cite{KL}, the invariants of the type of eqn.~\rf{Inv} do not depend on the choice of  coordinates and on the choice of the Weyl gauge. Thus any change of the Weyl gauge does not help to remove the cosmological singularity in the models studied in  \cite{Bars:2011th,Bars:2011aa,Bars:2012fq,Bars:2013qna}.

\end{document}